\documentclass[utf8]{frontiersSCNS} 

\usepackage{url,hyperref,lineno,microtype,subcaption}
\usepackage[onehalfspacing]{setspace}
\usepackage{color}
\usepackage{textcomp}

\def\keyFont{\fontsize{8}{11}\helveticabold }
\def\firstAuthorLast{Spiwok {et~al.}} 
\def\Authors{Vojtěch Spiwok\,$^{1,*}$, Martin Kurečka\,$^{2}$ and Aleš Křenek\,$^{2}$}
\def\Address{$^{1}$Department of Biochemistry and Microbiology,
Faculty of Food and Biochemical Technology,
University of Chemistry and Technology, Prague, Czech Republic \\
$^{2}$Institute of Computer Science, Masaryk University, Brno, Czech Republic }
\def\corrAuthor{Corresponding Author}

\def\corrEmail{spiwokv@vscht.cz}

\begin{document}
\onecolumn
\firstpage{1}

\title[AlphaFold Collective Variable]{Collective Variable for Metadynamics Derived from
AlphaFold Output} 

\author[\firstAuthorLast ]{\Authors} 
\address{} 
\correspondance{} 

\extraAuth{}

\maketitle

\begin{abstract}
\section{}
\noindent
AlphaFold is a neural network-based tool for the prediction of 3D structures
of protein. In CASP14, a blind structure prediction challenge, it
performed significantly better than other competitors, making
it the best available structure prediction tool. One of the
outputs of AlphaFold is the probability profile of residue-residue
distances. This makes it possible to score any conformation of
the studied protein to express its compliance with the AlphaFold
model. Here we show how this score can be used to drive protein
folding simulation by metadynamics and parallel tempering
metadynamics. Using parallel tempering metadynamics, we simulated the
folding of a mini-protein Trp-cage and $\beta$ hairpin
and predicted their folding equilibria.
We see the potential of the
AlphaFold-based collective
variable in applications beyond structure prediction, such as
in structure refinement or prediction of the outcome of a mutation.

\tiny
 \keyFont{ \section{Keywords:} AlphaFold, protein folding, protein structure prediction, metadynamics,
 deep learning, free energy simulation, collective variable}
\end{abstract}

\section{Introduction}

The introduction of the AlphaFold tool, and especially its second version \citep{AF2}, represents
a significant improvement in the prediction of 3D structures of proteins. In
CASP14 structure prediction competition, it outperformed other competitors,
both in overall as well as local accuracy of predicted structures. It is
very likely that AlphaFold will change the field of experimental structural
biology, and this field will in the future focus on the function of proteins rather
than the 3D structure itself.

Nevertheless, there are still limitations of AlphaFold such as limited
capacity to predict the outcome of a point mutation, to predict structures
of complexes with small-molecule ligands, to model an induced fit, and other
limitations. This provides an opportunity for biomolecular simulations
and their hybrid applications together with AlphaFold data.

AlphaFold 1 \citep{AF1} was introduced in CASP13 blind structure prediction competition
in 2018. One of the key premises in structure prediction by AlphaFold 1 is that
coevolving residues are likely to be located close to each other in a 3D
structure of a protein \cite{coevol}. The input of AlphaFold 1 is an amino acid sequence
of the modeled protein. Next, homologous sequences are found in a database
of sequenced proteins and aligned to give a multiple-sequence alignment.
This alignment is converted to various coevolution features. Distributions of
distances between residues are predicted from these features using an artificial
neural network. This neural network is trained on proteins with known 3D
structures from Protein Data Bank (PDB) \citep{PDB} and multiple sequence alignments
with their sequenced homologs. Finally, the resulting distance distribution
is used to predict the 3D structure of the modeled protein. 

AlphaFold 2 has significantly improved accuracy compared to AlphaFold 1 by the
introduction of novel neural network architectures and the integration of separated
parts into a more compact neural network pipeline. Coevolution is modeled rather
implicitly in this version. Both versions work with inter-residue distance maps.
That is, AlphaFold 2 produces a tensor with dimension $N$x$N$x$M$, where $N$ is the number
of residues and $M$ is the number of distance bins. This tensor stores the probabilities
(expressed as logits) of a given residue pair being found at a given distance. The
tensor makes it possible to evaluate any conformation of the modeled protein to
estimate how much it fits the AlphaFold prediction. The level of this
fitness was used in this work to drive a simulation of the protein, explore various conformations, and predict their equilibrium probabilities. For this
purpose we used the metadynamics method \citep{MTD} and its combination with parallel tempering \citep{ptmtd}
in explicitly modeled water.
The concept was tested on two artificial
fast-folding mini-proteins tryptophan cage (Trp-cage) and $\beta$ hairpin.

\section{Methods}
\subsection{AlphaFold}
The structure of Trp-cage miniprotein (the construct TC5b)
was predicted by AlphaFold 2 (initial release). \citep{AF2}
The structure of $\beta$ hairpin (the 16-residue
C-terminal fragment of the G-B1 protein sequence GEWTYDDATKTFTVTE)
by AlphaFold 2 (version 2.1.1). Both models were
in excellent agreement with the experimentally determined
structures (PDB ID 1L2Y \citep{trpcage}
and 2GB1 \cite{hairpin3d}), even if all homologous
structures were excluded from the
experimentally determined set of structures used by the program (by the option
\verb|--max_template_date=1969-12-31|).

\subsection{Molecular dynamics simulation}
All simulations were performed by Gromacs 2021 \citep{Gromacs} patched with
Plumed 2.7.2 \citep{Plumed2} modified to introduce the AlphaFold collective variable.
Source code of those extensions is publicly available 
and it will be added to Plumed in the near future.
A~Docker image of Gromacs built with this Plumed extension
is available as \url{ljocha/gromacs:2021-3.3} at Dockerhub.
The image supports both, single and double precision, SSE2/AVX2\_265/AVX\_512 instruction sets, and GPU acceleration.
It can be converted to Singularity for use in HPC computing centers in a~straightforward way.

In Trp-cage simulations, the system contained
the mini-protein, 11,112 or 1,602 TIP3P water
molecules \citep{TIP3P} (for metadynamics and parallel tempering metadynamics, respectively)
and one chloride anion to neutralize the charge.
In $\beta$ hairpin simulations, the system contained
the mini-protein, 11,136 or 1,625 TIP3P water
molecules (for metadynamics and parallel tempering metadynamics, respectively) and three sodium cations to neutralize the charge.

Mini-proteins were modeled
using Amber99SB-ILDN force field \citep{AMBER99SB-ILDN}. The time step was set to 2 fs and all bonds
involving hydrogen atoms were
constrained by the LINCS algorithm \citep{lincs}. Electrostatic interactions were evaluated
using the particle-mesh Ewald method. \citep{PME} Parrinello-Bussi thermostat \citep{vrescale} and
Parrinello-Rahman barostat \citep{barostat} were used to maintain constant temperature
and pressure, respectively.

For metadynamics, the system was first optimized by the steepest descent method
followed by 100 ps simulation in the NPT ensemble (constant number of particles,
pressure, and temperature) at 300 K. For parallel tempering metadynamics, the system was
equilibrated by 100 ps simulation in the NPT ensemble at 300 K, followed by 100 ps
simulation in the NVT ensemble (constant number of particles, pressure and temperature)
at each temperature used in parallel tempering metadynamics (278, 287, 295, 303, 312,
321, 329, 338, 346, 355, 365, 375, 385, 396, 406, 416, 427, 437, 448, 459, 470, 482,
493, 505, 517, 528, 539, 551, 562, 573, 584 and 595 K).

\subsection{Metadynamics with AlphaFold Collective Variable}
Molecular dynamics simulation makes it possible to realistically model the evolution of
a molecular system. Due to its high computational cost, it can sample relatively
short time scales, typically nanoseconds or microseconds for
explicitly solvated proteins. This is usually not enough
to accurately predict the long-term distribution of states of the system studied.

Multiple methods have been introduced to address this problem. One group of methods
uses artificial forces or potentials to help the system cross energy barriers \citep{REVIEW}.
The system, instead of being stuck in a single local energy minimum, explores
multiple energy minima. Metadynamics \citep{MTD} achieves this by ``flooding'' energy minima by
a history-dependent bias potential.

Another group of methods uses elevated temperatures to cross energy barriers. This
is the basis of parallel tempering \citep{pt}. The system is simulated in multiple replicas at
different temperatures. These replicas can occasionally swap their coordinates based
on pre-defined criteria. As a result, sampling of states at the temperature of
interest is more efficient compared to conventional simulation. Parallel tempering
and metadynamics have been successfully combined into parallel tempering metadynamics
\citep{ptmtd}.

This bias potential in metadynamics and parallel tempering metadynamics
is defined as a function of collective variables (CVs). A collective variable is
a descriptor of the molecular system studied predefined by the user. It must be
a differentiable function of the atomic coordinates. Furthermore, its value should
reflect the state of the simulated system, including metastable states. In this
work, we introduce a novel AlphaFold-based CV,
and we use it as a sole CV or
together with $\alpha$-RMDS CV \citep{alpharmsd}.

\begin{figure}[h!]
\begin{center}
\includegraphics{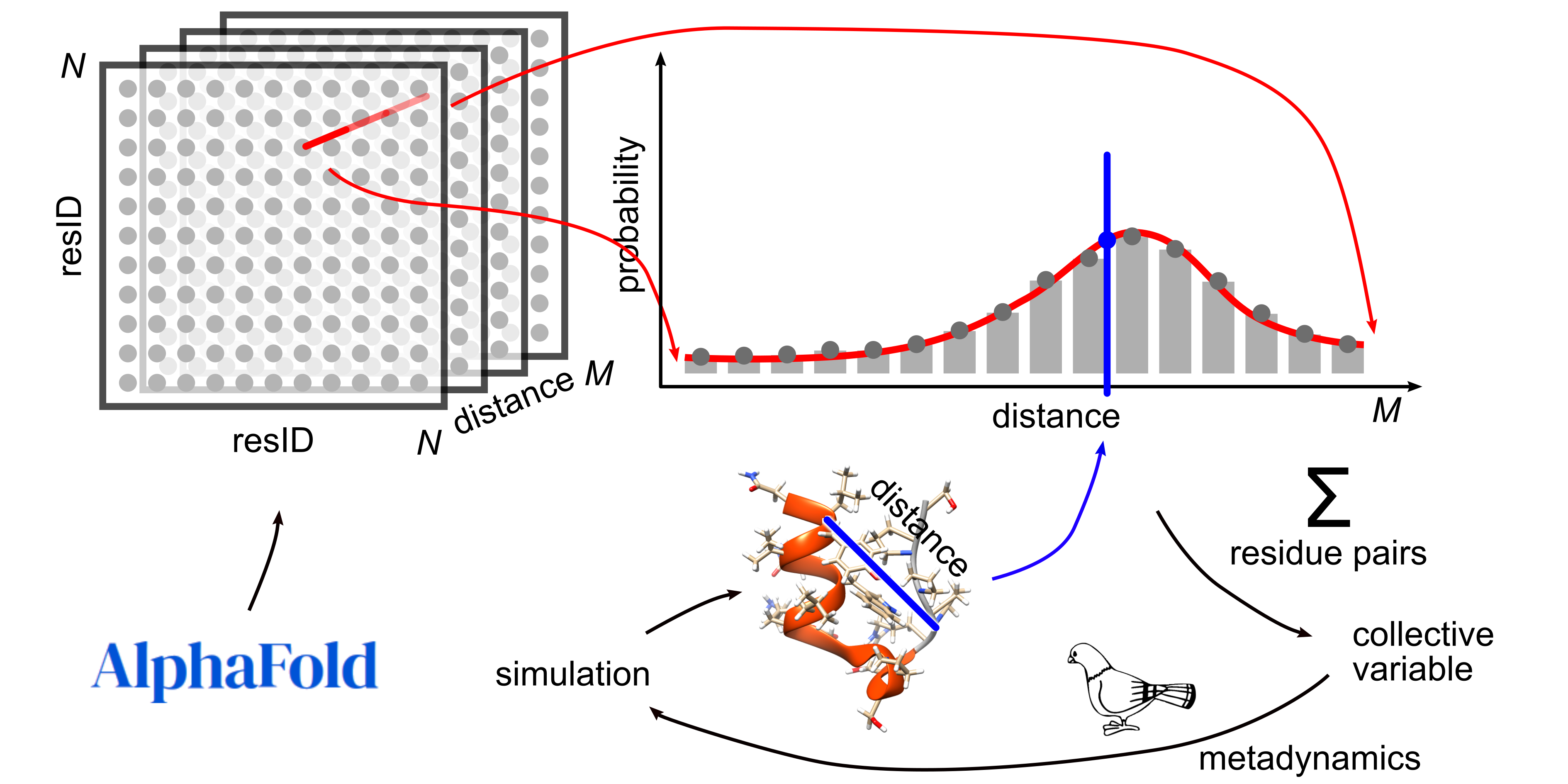}
\end{center}
\caption{Schematic representation of
AlphaFold-based CV.}\label{fig:scheme}
\end{figure}

Calculation of AlphaFold-based CV is presented in Figure \ref{fig:scheme}.
One of the outputs of AlphaFold 2 is a tensor~$D$ of distance probabilities with
dimension $N\times N\times M$, where $N$ is the number of residues
and $M$ is the number of distance bins (by default 64). For each residue
pair, this tensor was converted from logits to probabilities
$D[i,j,k]$ (where $i,j$ are the indexes of residues and $k$ is an index of bins)
and scaled to give the sum of probabilities along the $M$ axis equal to 1.

Given a~fixed conformation $C$ of the molecule, we can denote $\hat d_{i,j}$
the index of the distance bin where the distance of the residues $i,j$ falls.
Then the random variable $R_{i,j}$ with discrete values $0,1$ and the meaning
``for a~randomly chosen conformation, the distance between $i$ and $j$ falls into
the bin $\hat d_{i,j}$'' yields an expected value $ER_{i,j} = D[i,j,\hat d_{i,j}]$.
Hence, by linearity of expectation, the random variable
$R$ with the meaning ``number of inter-resudual distances of a~randomly chosen conformation 
that fall into the same bin as with $C$'' yields the expected value
\begin{equation}
ER = \sum_{i=1}^N\sum_{j=1}^{i-1} D[ i,j,\hat d_{i,j}].
\label{eq:er}
\end{equation}
Therefore, the sum can be interpreted as a~probabilistically based measure
to assess how AlphaFold would favor the conformation $C$.

In order to be used in metadynamics, the CV is desired to be smooth and must be differentiable with respect to atomic coordinates.
There are many interpolation techniques to choose from, we use an approach
derived from path collective variable definitions (path CV)~\citep{PathCV,propertymap},
simplified to the one-dimensional case.
Assuming $d$ to be an inter-residue distance between $i,j$ in our conformation~$C$, we calculate
\begin{equation}
P_{i,j}(d) = \frac{\sum_{k=1}^M D[i,j,k] e^{-\lambda (d-d_k)^2}}{\epsilon + \sum_{k=1}^M e^{-\lambda (d - d_k)^2}}
\label{eq:pahtcv}
\end{equation}
where $d_k$ is the inter-residue distance of $k$-th bin in $D$.
Other techniques (polynomial spline interpolation, etc.) could appear less complicated.
However, when considering the required differentiation, our approach is computationally efficient
(the exponential terms are reused) and less error-prone for implementation.

The value of $\lambda$ must be determined empirically -- low values make the curve smoother,
high values favor $P_i$s more strictly.
We used $\lambda=1,000 \text{ nm}^{-2}$ in our calculations.
The constant $\epsilon$ (pseudocount, not used in
the original path CVs) was introduced to improve
numerical stability when $d$ falls out of the $d_k$s range and Eq.~\ref{eq:pahtcv} approaches $0/0$. 
Finally, the values of $P_{i,j}(d)$ were calculated for all C$\alpha$-C$\alpha$
distances measured during the simulation and summed according to~Eq.~\ref{eq:er}. The
result was used as a collective variable in metadynamics simulations.

The AlphaFold output (the final model and the corresponding pickle file) was converted
by a Python script provided at GitHub
(\url{https://github.com/spiwokv/af2cv}).
It converts the 3D structure (in Gromacs format) and the pickle file into
a Plumed input by command:
\begin{verbatim}
python af2cv.py model.gro model.pkl > plumed.dat
\end{verbatim}
The resulting output (plumed.dat) must be modified for the given type of calculation,
e.g. for monitoring of the CV, for metadynamics or for other free energy modeling methods
available in Plumed.

In metadynamics we used either AlphaFold-based CV or its combination with $\alpha$-RMSD.
Metadynamics floods the free-energy minima by a bias potential composed of Gaussian hills.
The widths of these hills were 0.1 (Trp-cage) or
0.04 ($\beta$ hairpin)
in the direction of AlphaFold-based CV and 0.1
in the direction of $\alpha$-RMSD. Well-tempered metadynamics \citep{WT-MTD}, which reduces the heights
of hills with the progress of the simulation, was used. The height of the hills was set to
0.5 kJ/mol and the bias factor was set to 8.

Parallel tempering metadynamics \citep{ptmtd} also used either
AlphaFold-based CV
or its combination with $\alpha$-RMSD. The widths and heights of the hills and the bias factor were the same
as in metadynamics. Replica exchange attempts were
made every 1 ps. Free energy surfaces were calculated using Metadynminer \citep{metadynminer}.
3D structures were visualized by UCSF Chimera \citep{chimera}.

\begin{figure}[h!]
\begin{center}
\includegraphics{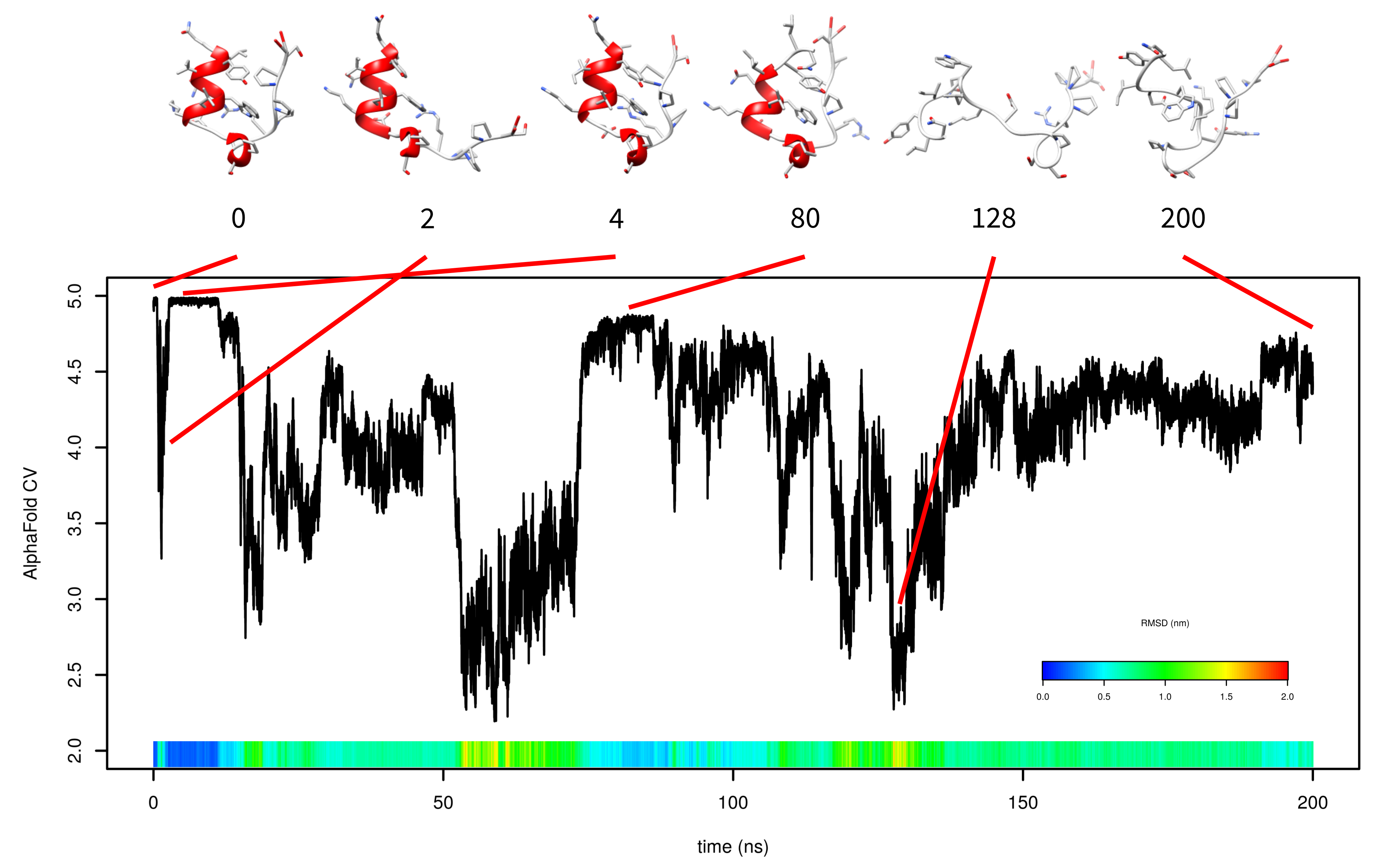}
\end{center}
\caption{Evolution of AlphaFold-based
         CV in metadynamics
         simulation of Trp-cage
         with AlphaFold-based
         CV with selected frames.
         RMSD is depicted in color scale.}
         \label{fig:mtd1}
\end{figure}

\section{Results}
\subsection{Trp-cage}

First, 200 ns metadynamics was performed only with
AlphaFold-based CV (Figure \ref{fig:mtd1}).
The simulation started from the folded state. The folded state
corresponds to AlphaFold-based CV
values between 4.9 to 5. After approximately
600 ps it partially unfolded and AlphaFold-based
CV dropped to approximately 3.
This unfolding consisted of detachment of the C-terminal tail
from the N-terminal $\alpha$-helix. The C-terminal tail returned at time 2.6 ns and the structure returned to the folded state.
It stayed there up to 11 ns. This is represented
in Figure \ref{fig:mtd1} by snapshots at 0, 2 and 4 ns.

At the time of approximately 80 ns, the system
returned to near-native state
with the AlphaFold-based
CV between 4.80 and 4.85. It differed from the native
state by unwinding of the three N-terminal residues from the
$\alpha$-helix and a slightly different position of the C-terminus.
From the start to time 80 ns, there was some content of
$\alpha$-helix in the structure. After that, the protein lost any
helix content and was not able to fold.

Unfortunately, the free energy
surface calculated by this metadynamics simulation was not realistic
(data not shown) because the unfolded state was calculated as
significantly more stable than the folded state. This was due to
lack of folding events.

\begin{figure}[h!]
\begin{center}
\includegraphics{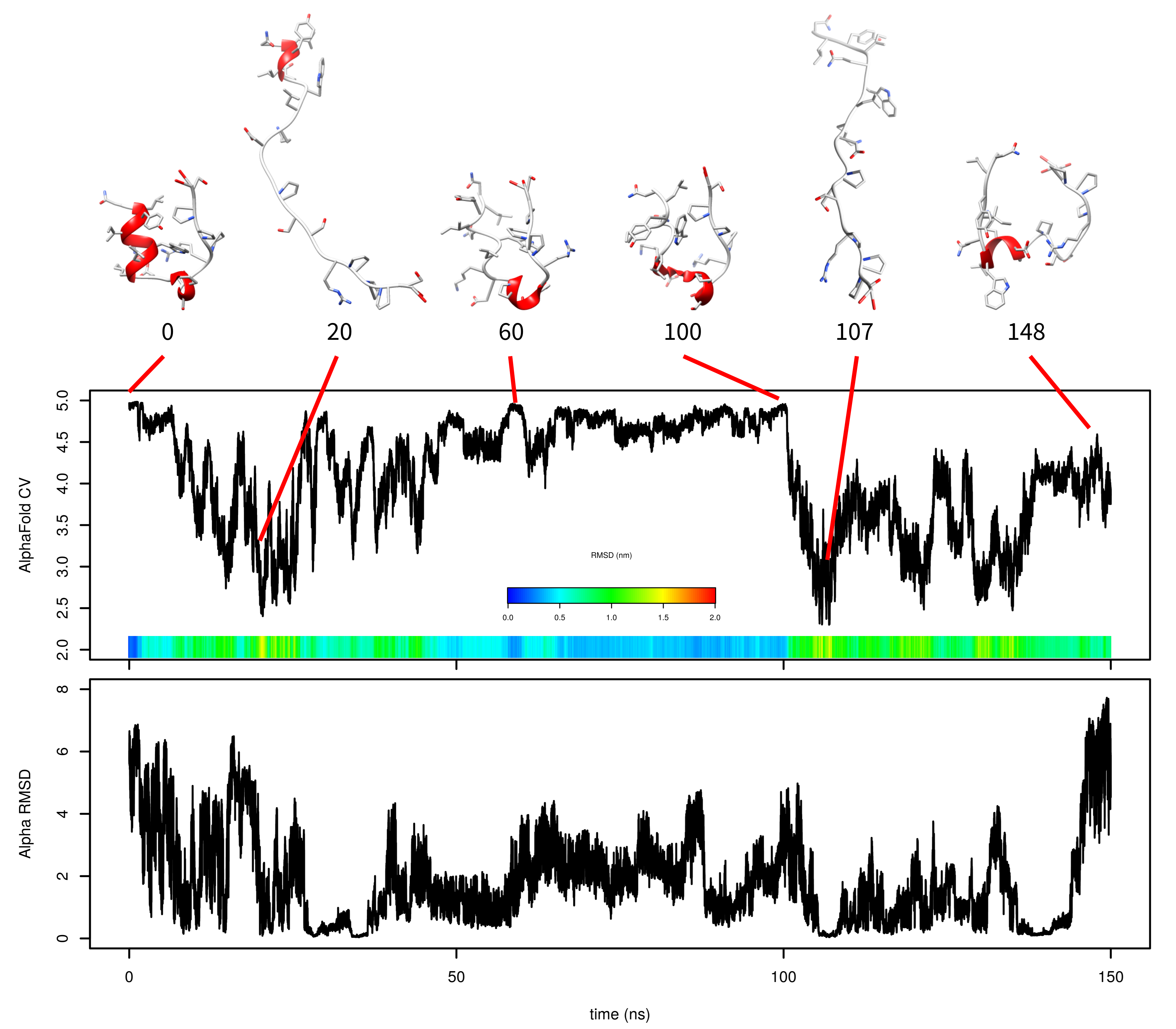}
\end{center}
\caption{Evolution of AlphaFold-based
         CV and $\alpha$-RMSD in metadynamics
         simulation of Trp-cage
         with both CVs with selected frames.
         RMSD is depicted in color scale.}
         \label{fig:mtd2}
\end{figure}

The simulation with one AlphaFold-based
CV indicated that the formation
of $\alpha$-helix is critical and that
AlphaFold-based CV itself cannot
efficiently accelerate it. For this reason, we added a second
CV ($\alpha$-RMSD) to accelerate helix formation. The results of
this 150 ns simulation are depicted in Figure \ref{fig:mtd2}.

Similarly to the previous simulation, there was relatively fast unfolding.
The system explored structures with high values of
AlphaFold-based CV
between 60 and 100 ns. The value of
AlphaFold-based CV was fluctuating
between 4.4 and 4.96 at this stage. These structures were
very similar to the native state in overall shape, but the
N-terminal $\alpha$-helix was not formed.

The simulation
also explored states with higher $\alpha$-RMSD CV, which is
depicted in Figure \ref{fig:mtd2} at 148 ns. This structure
is characterized by a helix-like structure of residues 6-13
(Figure \ref{fig:mtd2} shows residues 6-9 as helix, because
the rest do not meet the definition of $\alpha$-helix
used by UCSF Chimera).

Again, the predicted free energy surface was not realistic
(the unfolded state was significantly more favored than the folded
one; data not shown). This can be explained by the lack of folding
events.

Since it was not possible to observe enough folding events to
accurately predict the folding free energy surface, we replaced
metadynamics with parallel tempering metadynamics. The combination
of metadynamics with parallel tempering makes it possible to
accelerate degrees of freedom that are not covered by CVs.

\begin{figure}[h!]
\begin{center}
\includegraphics{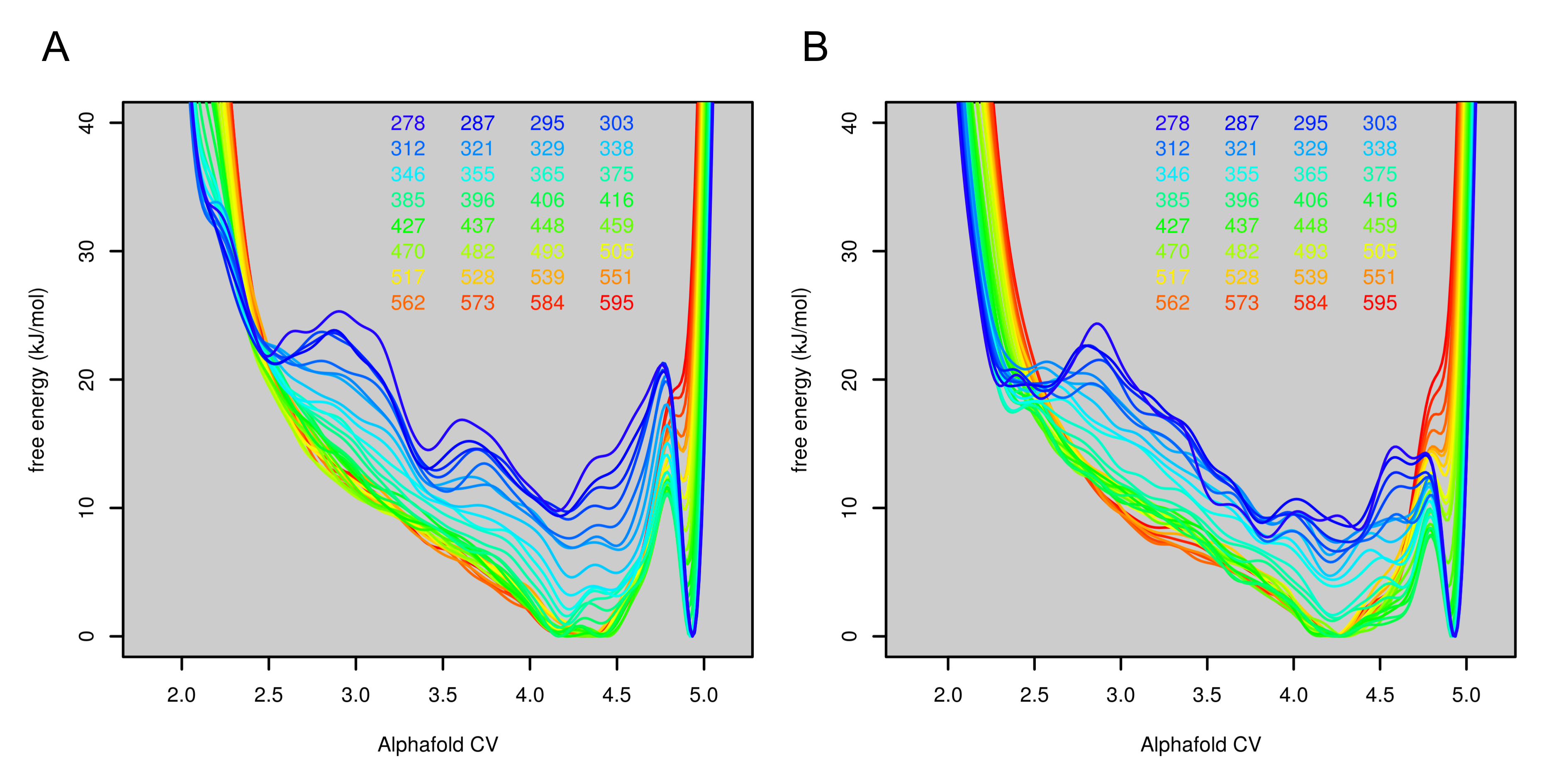}
\end{center}
\caption{Free energy surfaces of Trp-cage
         (as functions of AlphaFold-based
         CV) calculated at different temperatures
         in the run stared from folded
         (\textbf{A}, the first run)
         and unfolded (\textbf{B}, the second run)
         state.}\label{fig:ptmtdfes}
\end{figure}

The first parallel tempering metadynamic simulation started from
the native structure of the protein. Before the parallel tempering
metadynamics simulation, each replica was equilibrated
by 100 ps molecular dynamics simulation at the corresponding
temperature. This was usually not long enough to unfold the protein,
so most simulations started from the native or near-native structure.
The predicted free energy surfaces at different
temperatures are depicted in Figure \ref{fig:ptmtdfes}A.

The folded state is modeled as the global minimum for lower
temperatures (up to 375 K, 102 \textdegree C). Above this the 
unfolded state was more favored. The conversion of the free
energy surface to probability (as $\exp(-G/kT)$) and integration of
probability of AlphaFold score higher and lower than
4.75 (estimated border between the unfolded and folded
state) revealed that the protein is stable in its native structure
up to 329 K (56 \textdegree C). This was in good agreement with
the experimentally determined melting temperature
(42 \textdegree C).

The fact that the simulation started mostly from the folded state
may bias the free energy surface towards the folded state.
To rule out this possibility, we performed another simulation
(second run) starting from the first run after 10 ns. At this
point, all replicas were unfolded (only replica 18 was in a
state similar to the structure at 2 ns in Figure \ref{fig:mtd1}).
All other settings were the same as in the first run. The free energy
surface is depicted in Figure \ref{fig:ptmtdfes}B. Free energy
surfaces from the first and the second run are very similar.
The only notable difference was in the lower barrier between the
folded and unfolded states.

Free energy surfaces calculated by parallel tempering metadynamics
may be stable thanks to transitions between different states of the
system, but also as an artifact of replica exchanges.
Imagine a parallel tempering (or parallel
tempering metadynamics) with just two replicas, one starting from
a folded and one from an unfolded protein. High number of replica
exchanges causes the single temperature trajectory
switches between the folded and
unfolded states, even in the absence of any real
folding and unfolding
events. The results of such simulations may be wrongly interpreted
as an equilibrium between the folded and unfolded state.

\begin{figure}[h!]
\begin{center}
\includegraphics{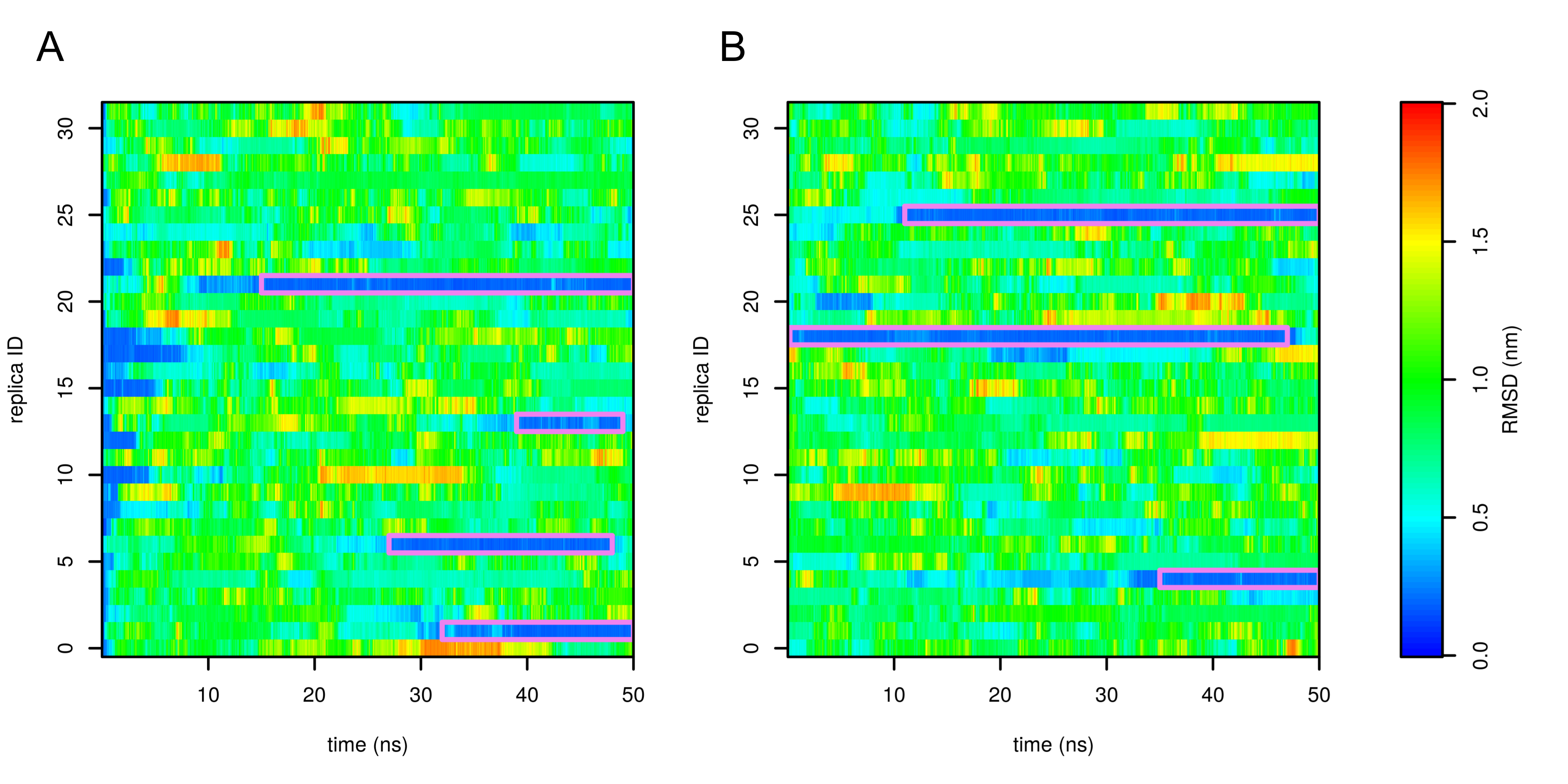}
\end{center}
\caption{Profiles of RMSD
         of Trp-cage
         as a function of time calculated
         for demultiplexed trajectories
         from the run stared from folded
         (\textbf{A}, the first run)
         and unfolded (\textbf{B}, the second run)
         state. The folded state is highlighted by
         a purple frame.}\label{fig:demuxed}
\end{figure}

To avoid this artifact, we performed a demultiplexing (``demuxing'')
of replicas to obtain continuous trajectories, regardless of the 
evolution of temperature. The evolution of root-mean-square
deviation (RMSD) from the native structure in demuxed trajectories
is depicted in Figure \ref{fig:demuxed}. In the first and the second
runs, we observed four and three folding events, respectively. In
general, there were typically two replicas in the folded state.
This indicates that the folding free energy surfaces in
Figure \ref{fig:ptmtdfes} are stable and are not affected by either the starting state or the number of folding events.

\begin{figure}[h!]
\begin{center}
\includegraphics{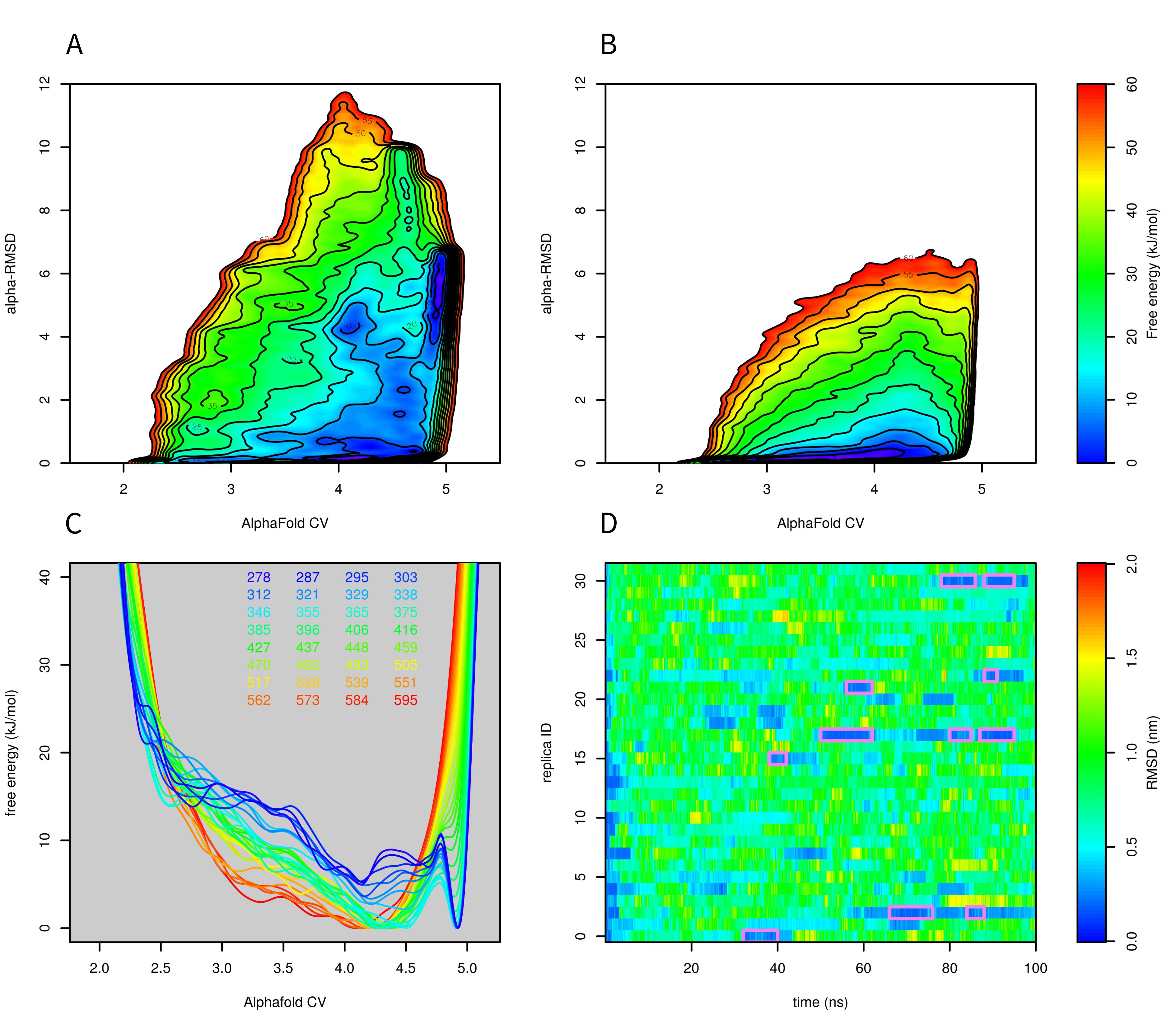}
\end{center}
\caption{Results of parallel tempering metadynamics
         simulation of Trp-cage with
         AlphaFold-based CV and
         $\alpha$-RMSD collective variables.
         \textbf{A} - free energy surface at 303 K,
         \textbf{B} - free energy surface at 595 K,
         \textbf{C} - free energy surfaces calculated at
         different temperatures and, 
         \textbf{D} - profiles of RMSD as a function of 
         time calculated for demultiplexed trajectories.}\label{fig:ptmtd3}
\end{figure}

Finally, we tested parallel tempering metadynamics with two collective variables,
AlphaFold-based CV and $\alpha$-RMSD.
It was necessary to prolong the simulation from 50 to
100 ns. The results are depicted in Figure \ref{fig:ptmtd3}. Figure \ref{fig:ptmtd3}A and B
compares the free energy surface at 303 and 595 K. The former is characterized by two main
minima. The minimum corresponding to the folded structure
is located at AlphaFold-based CV
$\sim$ 5 and $\alpha$-RMSD $\sim$ 6. The unfolded minimum is at the bottom of the plot.
There are several other notable local minima, namely, at high values of $\alpha$-RMSD, which
corresponds to a structure with a long helix, or at
AlphaFold-based CV $\sim$ 4 and
$\alpha$-RMSD $\sim$ 4, which corresponds to the structure with the N-terminal helix formed
but the C-terminal tail detached from the helix. At 595 K, the unfolded minimum was
the only minimum of the system. Figure \ref{fig:ptmtd3}C and D are in good agreement
with the results of parallel tempering simulations with
AlphaFold-based CV as the only
collective variable. A reasonable number of folding events was observed.

\begin{figure}[h!]
\begin{center}
\includegraphics{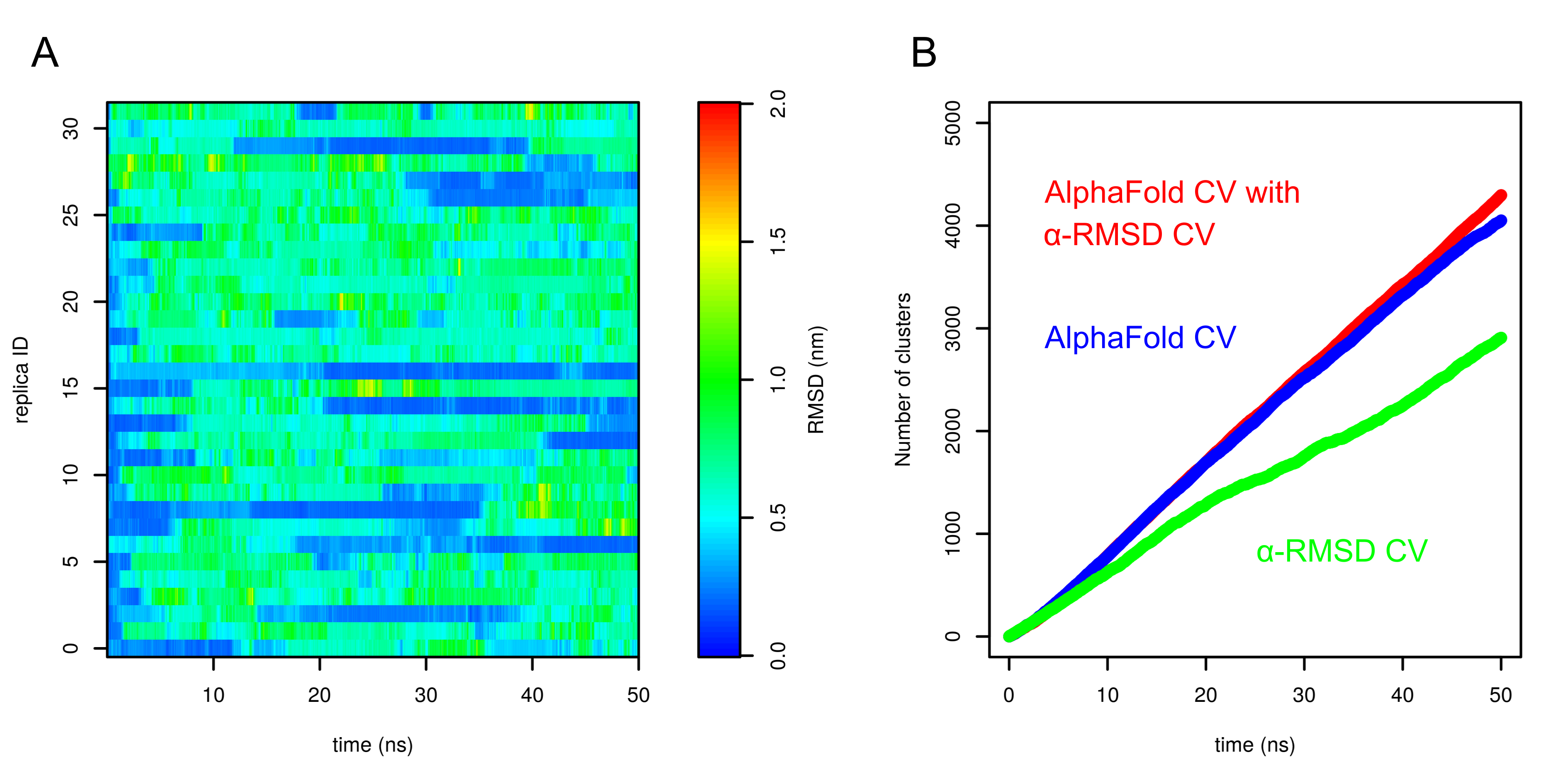}
\end{center}
\caption{Comparison of parallel tempering metadynamics
         simulation of Trp-cage with
         $\alpha$-RMSD CV with other CVs.
         \textbf{A} - profiles of RMSD in 
         parallel tempering metadynamics with
         $\alpha$-RMSD CV as a function of 
         time calculated for demultiplexed trajectories and,
         \textbf{B} - comparison of the cumulative number
         of conformational clusters explored by parallel
         tempering metadynamics with different
         CVs.}\label{fig:alpharmsd}
\end{figure}

In order to compare the performance of
AlphaFold-based CV and its combination
with $\alpha$-RMSD we carried out a parallel tempering simulation with
$\alpha$-RMSD as the only CV. Other parameters were the same as in
simulations with the AlphaFold-based CV. Surprisingly, we
observed multiple
folding events (Figure \ref{fig:alpharmsd}A).
However, it is apparent that
unfolded structures in these simulations do not significantly divert
from the native structure. In other words, the sampling of various 
conformation states is much more intensive in the simulation with
AlphaFold-based CV in combination of AlphaFold and $\alpha$-RMSD CVs,
compared to $\alpha$-RMSD CVs.

This was further demonstrated by assessment of the number of
conformational clusters explored in the simulation. Structures
sampled in each parallel tempering simulation at 303 K were
analyzed by the clustering method by Daura and co-workers \cite{daura}
(cut-off set to 0.1 nm).
Figure \ref{fig:alpharmsd}B shows the evolution of the cumulative
number of clusters. Clearly, the number of clusters sampled by 
$\alpha$-RMSD CV is significantly lower than for
AlphaFold-based CV.
The combination of AlphaFold and $\alpha$-RMSD CVs gives little extra
sampling.

\begin{figure}[h!]
\begin{center}
\includegraphics{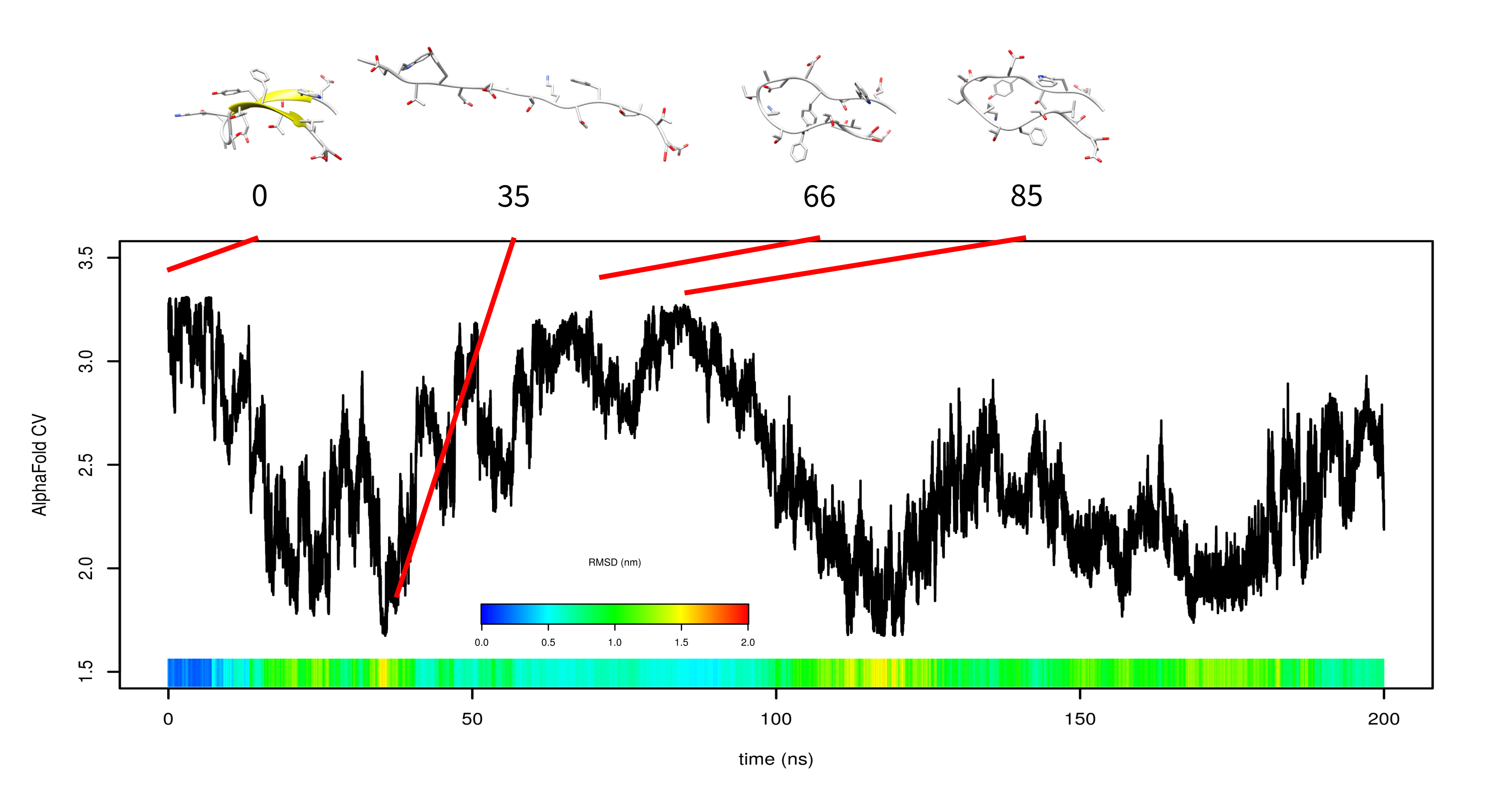}
\end{center}
\caption{Results of metadynamics simulation
         of $\beta$-hairpin with
         AlphaFold collective variable
         with selected structures.
         RMSD is depicted in color scale.}\label{fig:hairpin1}
\end{figure}

\subsection{$\beta$ Hairpin}
Formation of $\beta$-sheet is in general slower than
formation of $\alpha$-helix \cite{helixvssheet}.
Furthermore, it is more
difficult to accelerate it. To evaluate the performance of
AlphaFold\textcolor{red}{-based} CV on the formation
of $\beta$-sheet structures
we studied folding of a model $\beta$ hairpin mini-protein,
which is composed of a single anti-parallel $\beta$-sheet.
Similarly to Trp-cage metadynamics
simulation, after complete unfolding at
the beginning of 200 ns metadynamicssimulation,
the system explored structures
with high AlphaFold\textcolor{red}{-based} CV (at time
66 or 85 ns, Figure \ref{fig:hairpin1}).
These structures were spatially similar to the native
structure, however, the secondary structure
was formed incorrectly.

\begin{figure}[h!]
\begin{center}
\includegraphics{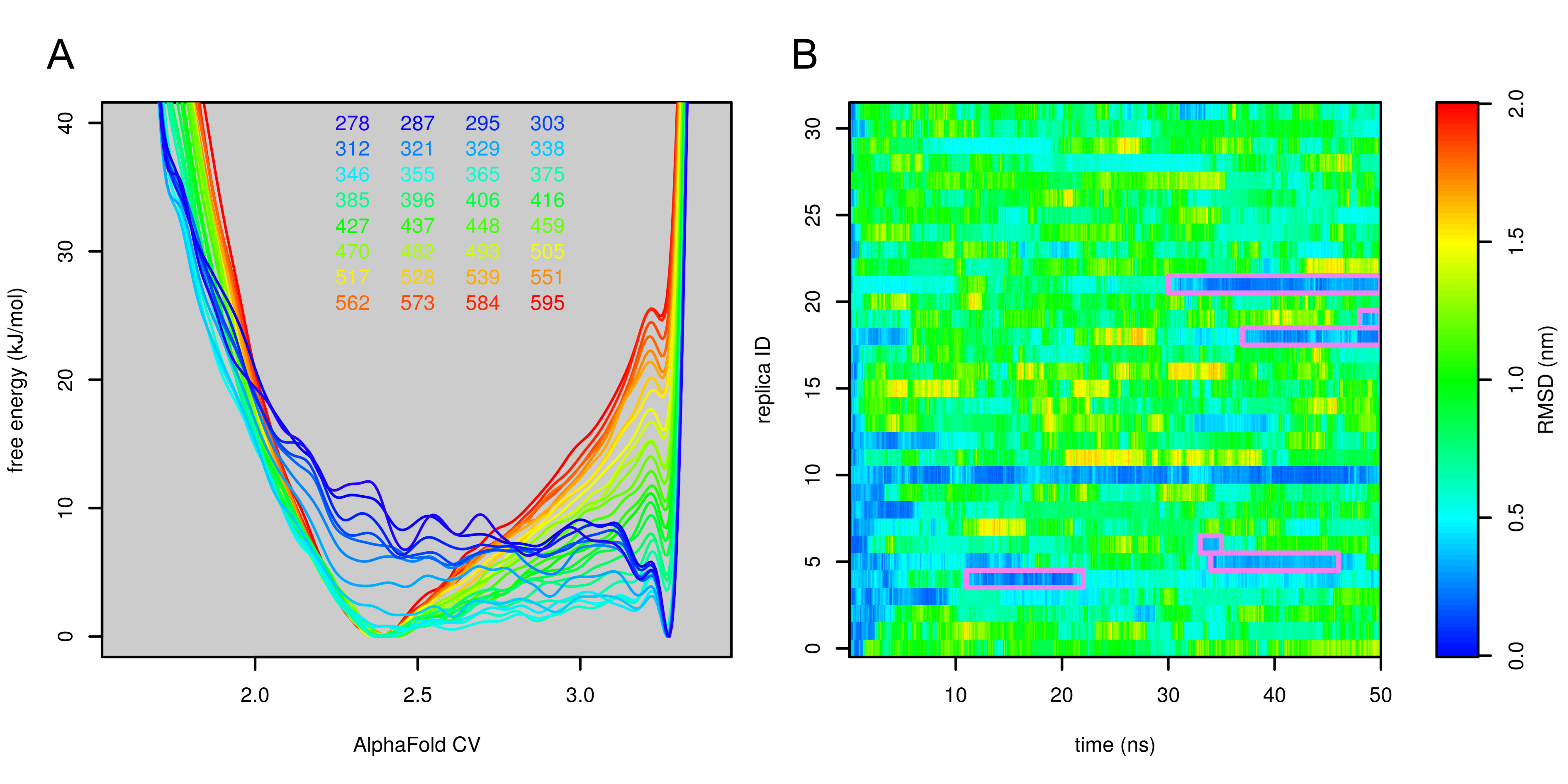}
\end{center}
\caption{Results of parallel tempering metadynamics
         simulation of $\beta$-hairpin with
         AlphaFold collective variables.
         \textbf{A} - Free energy surfaces calculated
         at different temperatures and,
         \textbf{B} - profiles of RMSD as a function of 
         time calculated for demultiplexed trajectories.}\label{fig:hairpin2}
\end{figure}

Similarly to Trp-cage, combination of parallel
tempering with metadynamics made it possible to accurately predict
its free energy surface at different temperatures and to observe
multiple folding events (Figure \ref{fig:hairpin2}). $\beta$ Hairpin
was predicted to prefer the folded state at low temperatures and
the unfolded state at higher temperatures. The melting temperature was
estimated as 287 K. This is in good agreement with
the experimentally determined fraction of the native structure
at 278 K, which is 40 \% \cite{hairpintm}.

\section{Discussion}

The performance of metadynamics driven by AlphaFold-based
CV can be assessed by comparison with
unbiased simulations, parallel tempering simulations, and metadynamics using other CVs.
One of model 
mini-proteins used in this study -- Trp-cage -- folds with the mean folding time
equal to 14 $\mu$s in a simulation with a similar setup \citep{deshaw}. In a parallel tempering
simulation (without metadynamics) with a similar setup and the same duration (200 ns) as
in our previous study, we did not observe any folding events \citep{anncolvar}. With the
neural-network-approximated solvent-accessible surface area as a CV we observed more
folding events than in this study (eight compared to four or three in this study). Both
studies used 200 ns parallel tempering metadynamics. However, for approximation of the
solvent-accessible surface area by a neural network, it is necessary to have a series of
folded and unfolded structures of the system. We obtained these from the 208 $\mu$s trajectory
kindly provided by D.E. Shaw Research. In contrast,
AlphaFold-based CV can be built
just using the sequence of a protein. Therefore, it does not suffer from
the ``chicken and egg''
problem of the necessity of using folding trajectories to simulate folding.

Our results have shown that the critical process in the folding
of Trp-cage accelerated by AlphaFold-based
CV is the formation of the
secondary structure. Without this, metadynamics can force formation
of structures similar to the native one, but wrong in terms of the
secondary structure. This cannot be easily solved by the
$\alpha$-RMSD CV. Replacement of metadynamics by parallel tempering
metadynamics helped to solve this problem.
$\alpha$-RMSD CV itself makes it possible to fold
and unfold Trp-cage in a reasonable time scale; however, a
significantly lower number of conformations is sampled.

AlphaFold-based CV was also tested on $\beta$ hairpin,
which was used as a representative of $\beta$-sheet mini-protein.
Similarly to Trp-cage, AlphaFold-based CV was not able to fold
the mini-protein in 200-ns metadynamics, but was able to fold
it and predict the temperature-dependent free energy surface
by parallel tempering metadynamics.

We used the concept of path CVs (Equation \ref{eq:pahtcv}) \citep{PathCV} to convert
the discrete distance probability profile into a continuous one. The path CV includes
a prefactor $\lambda$ that must be set prior the application of the CV. Here we set
$\lambda$ equal to 1,000 nm\textsc{$-2$}. This value was chosen based on the plot of the
sample probability profiles (data not shown). We believe that the same value of
$\lambda$ can be used in studies of other proteins, because the distance values,
for which these profiles are constructed, are the same or very similar (the
Supplementary information of the reference \citep{AF2} states ``The bins cover the range
from 2 \r{A} to 22 \r{A}'').

The parameter $\epsilon$ used in Equation \ref{eq:pahtcv} has not been used, to our knowledge, in the context of path CV. It can be used to make the simulation
more stable by elimination of high gradient
when the numerator and the denominator of Equation
\ref{eq:pahtcv} are close to zero. In this study, we used this equation to approximate
a probability. In general applications of path CVs, this coefficient may cause artifacts
and must be used carefully.

In parallel tempering and parallel tempering
metadynamics it is necessary to keep the size of
a simulation box small. This is because the potential
energy distribution in large systems (large boxes
with large numbers of water molecules) is relatively
narrow as a result of an averaging effect. This causes
the overlap of potential energy histograms of two
neighboring replicas to be small, which causes a low
probability of coordinate exchange and thus poor
performance of parallel tempering. On the other
hand, small box size increases the risk of artifacts
caused by interactions of the simulated protein
with its replicas from the neighboring periodic
boxes. The trajectories from parallel tempering
metadynamics simulations were visually inspected
(data not shown). This revealed that self-interactions
are relatively rare and are limited to head-to-tail
interactions of fully unfolded proteins. Therefore, we
believe that self-interactions do not cause any
significant artifacts.

Similar approach as presented here was applied
by Nassar and coworkers \cite{nassar}. Using the output from
other machine-learning-based protein structure modeling tools,
they folded multiple significantly larger proteins, however,
in substantially longer simulations.

In principle, it is possible to use AlphaFold to predict
the native structure of the studied protein and then
use RMSD from the predicted native structure as a collective
variable (here $R^2$ of RMSD from the native structure and the
AlphaFold CV is 0.83 for Trp-cage). We see two major
differences between the RMSD from the native structure and
AlphaFold CV. First, bell-shaped distance probability
profiles have different widths for different residue
pairs. Narrow profiles give rise to higher energy
gradients; thus, they have higher priority than wide
profiles. We believe that this prioritization of residue
pairs may play a role in CV performance.

Second, AlphaFold-based CV is an example of a ``soft''
collective variable. Recall that the AlphaFold-based CV represents
the expected number of pairs of residues whose distance match
the current structure. Thus, it seems very likely that the local
minima in a fixed basis of attraction could have similar
AlphaFold-based CV values. If true, the AlphaFold-based CV
would change only if the simulation left the initial basis
of attraction. On the contrary, the RMSD of the native structure
is a variable that increases continuously with the divergence
from the native structure. Hence, it can vary much even when
restricted to minima on a fixed basis of attraction. This
could cause AlphaFold-based CV to provide faster divergence
from the initial structure.

A similar behaviour could be expected from other collective
variables that do not utilize the native structure, such as
$\alpha$-RMSD. As our experimental comparison shows, $\alpha$-RMSD
was capable of guiding the simulations into several folding
events; however, it could not lead it into a very divergent
basis of attraction. In contrast, the AlphaFold-based CV seems to
have similar values on similar bases of attraction causing
the simulation to move faster toward very different configurations.
The very same property, of course, prevents it from convergence
to minima in different bases of attraction; hence a second
CV is needed to ensure the convergence.

In the future, we can imagine a more focused version of
AlphaFold-based CV.
Since it is not the purpose of an AlphaFold-CV-driven simulation
to predict the structure of a protein, which can be done much
more efficiently by AlphaFold itself, we see its application in
refinement focused on protein loops, domains, pockets, 
for example to accelerate induced fit
in docking simulations.
Instead of calculating the sum of $P(d)$ across all residues,
the sum can be calculated on a predefined set of residues.
This would make it possible to focus
AlphaFold-based CV to certain part of the protein.
Analogously, it would be possible to split
the AlphaFold-based CV into two CVs,
one focused locally (that is, on the secondary structure)
and one globally.

\section*{Conflict of Interest Statement}
The authors declare that the investigation was carried out
in the absence of any commercial or financial relationships
that could be construed as a potential conflict of interest.

\section*{Author Contributions}
All authors developed AlphaFold-based CV.,
M.K. and A.K. prepared the software, V.S. carried out the simulations,
and all authors analyzed the results and wrote the
manuscript.

\section*{Funding}
This study was funded by Czech Science Foundation (22-29667S).
Computational resources were provided by
the Ministry of Education, Youth and Sports of the Czech Republic
by the CESNET (LM2018140), the CERIT Scientific Cloud (LM2015085)
and ELIXIR-CZ project (LM2018131), provided under the program
``Projects of Large Research, Development, and Innovations
Infrastructures''.

\section*{Acknowledgments}
Authors would like to thank DeepMind Technologies Limited
for making AlphaFold available to the scientific community.

\section*{Supplemental Data}
Supplementary Material was not provided.

\section*{Data Availability Statement}
Input files are available via Plumed-Nest
(\href{https://www.plumed-nest.org/eggs/22/005/}
{plumID:22.005}) \citep{nest}
and Zenodo (DOI:
\href{https://doi.org/10.5281/zenodo.5960030}
{10.5281/zenodo.5960030}).
Source code of the Plumed extension is available at
\url{https://github.com/kurecka/plumed2},
recipes for building Gromacs in a~Docker container with these extensions are available at
\url{https://github.com/ljocha/gromacs-plumed-docker}.

\bibliographystyle{frontiersinSCNS_ENG_HUMS}
\bibliography{test}
\end{document}